\documentclass[%
 reprint,
superscriptaddress,
preprintnumbers,
nofootinbib,
 showpacs, amsmath,amssymb,
 aps,
prl,
floatfix,
]{revtex4-1}

\usepackage[colorlinks = true,
            linkcolor = blue,
            urlcolor  = blue,
            citecolor = blue,
            anchorcolor = blue]{hyperref}
            
\usepackage{placeins}

\usepackage{graphicx}
\usepackage{dcolumn}
\usepackage{bm}
\usepackage{hyperref}

\usepackage{booktabs} 
\AtBeginDocument{
\heavyrulewidth=.08em
\lightrulewidth=.05em
\cmidrulewidth=.03em
\belowrulesep=.65ex
\belowbottomsep=0pt
\aboverulesep=.4ex
\abovetopsep=0pt
\cmidrulesep=\doublerulesep
\cmidrulekern=.5em
\defaultaddspace=.5em
}

\usepackage{textcomp}

\usepackage{siunitx}
\DeclareSIUnit{\litre}{\ell}

\begin{document}

\preprint{FERMILAB-PUB-23-625-PPD}

\title{First Results from a Broadband Search for Dark Photon Dark Matter in the 44\,to\,52\,$\mu$eV Range with a Coaxial Dish Antenna}


\author{Stefan Knirck}\email[Correspondence to: knirck@fnal.gov]{}
\affiliation{Fermi National Accelerator Laboratory, Batavia, IL 60510, USA}

\author{Gabe Hoshino}
\affiliation{Department of Physics, University of Chicago, Chicago, IL 60637, USA}


\author{Mohamed H. Awida}
\affiliation{Fermi National Accelerator Laboratory, Batavia, IL 60510, USA}

\author{Gustavo I. Cancelo}
\affiliation{Fermi National Accelerator Laboratory, Batavia, IL 60510, USA}

\author{Martin Di Federico}
\affiliation{Fermi National Accelerator Laboratory, Batavia, IL 60510, USA}
\affiliation{Universidad Nacional del Sur, IIIE-CONICET, Argentina}

\author{Benjamin~Knepper}
\affiliation{Fermi National Accelerator Laboratory, Batavia, IL 60510, USA}
\affiliation{Enrico Fermi Institute, University of Chicago, Chicago, IL 60637, USA}

\author{Alex Lapuente} 
\affiliation{Department of Physics, University of Chicago, Chicago, IL 60637, USA}

\author{Mira Littmann} 
\affiliation{Department of Physics, University of Chicago, Chicago, IL 60637, USA}

\author{David W. Miller} 
\affiliation{Department of Physics, University of Chicago, Chicago, IL 60637, USA}
\affiliation{Enrico Fermi Institute, University of Chicago, Chicago, IL 60637, USA}
\affiliation{Kavli Institute for Cosmological Physics, University of Chicago, Chicago IL 60637, USA}

\author{Donald V. Mitchell} 
\affiliation{Fermi National Accelerator Laboratory, Batavia, IL 60510, USA}

\author{Derrick Rodriguez} 
\affiliation{Department of Physics, University of Chicago, Chicago, IL 60637, USA}

\author{Mark K. Ruschman} 
\affiliation{Fermi National Accelerator Laboratory, Batavia, IL 60510, USA}

\author{Matthew A. Sawtell} 
\affiliation{Fermi National Accelerator Laboratory, Batavia, IL 60510, USA}

\author{Leandro Stefanazzi} 
\affiliation{Fermi National Accelerator Laboratory, Batavia, IL 60510, USA}

\author{Andrew Sonnenschein} 
\affiliation{Fermi National Accelerator Laboratory, Batavia, IL 60510, USA}
\affiliation{Enrico Fermi Institute, University of Chicago, Chicago, IL 60637, USA}

\author{Gary W. Teafoe} 
\affiliation{Fermi National Accelerator Laboratory, Batavia, IL 60510, USA}


\author{Daniel Bowring}
\affiliation{Fermi National Accelerator Laboratory, Batavia, IL 60510, USA}

\author{G. Carosi}
\affiliation{Lawrence Livermore National Laboratory, Livermore, California 94550, USA}

\author{Aaron Chou}
\affiliation{Fermi National Accelerator Laboratory, Batavia, IL 60510, USA}

\author{Clarence L. Chang}
\affiliation{Argonne National Laboratory, Lemont, IL 60439, USA}
\affiliation{Department of Astronomy \& Astrophysics, University of Chicago, Chicago, IL 60637, USA}
\affiliation{Kavli Institute for Cosmological Physics, University of Chicago, Chicago IL 60637, USA}

\author{Kristin Dona}
\affiliation{Department of Physics, University of Chicago, Chicago, IL 60637, USA}

\author{Rakshya~Khatiwada}
\affiliation{Fermi National Accelerator Laboratory, Batavia, IL 60510, USA}
\affiliation{Department of Physics, Illinois Institute of Technology, Chicago, IL 60616, USA}

\author{Noah A. Kurinsky}
  \affiliation{SLAC National Accelerator Laboratory/Kavli Institute for Particle Astrophysics and Cosmology, Menlo Park, Stanford University, Stanford, CA 94025, USA}

\author{Jesse Liu}
\affiliation{Cavendish Laboratory, University of Cambridge, Cambridge CB3 0HE, UK}

  \author{Cristián Pena}
\affiliation{Fermi National Accelerator Laboratory, Batavia, IL 60510, USA}

\author{Chiara P. Salemi}
  \affiliation{SLAC National Accelerator Laboratory/Kavli Institute for Particle Astrophysics and Cosmology, Menlo Park, Stanford University, Stanford, CA 94025, USA}

\author{Christina W. Wang}
  \affiliation{California Institute of Technology, Pasadena, CA 91125, USA}

\author{Jialin Yu}
\affiliation{Department of Physics, Illinois Institute of Technology, Chicago, IL 60616, USA}

\collaboration{BREAD}

\date{\today}

\begin{abstract}
We present first results from a dark photon dark matter search in the mass range from 44~to~52\,$\mu{\rm eV}$ ($10.7 - 12.5\,{\rm GHz}$) using a room-temperature dish antenna setup called GigaBREAD.
Dark photon dark matter converts to ordinary photons on a cylindrical metallic emission surface with area \SI{0.5}{\square\metre} and is focused by a novel parabolic reflector onto a horn antenna. Signals are read out with a low-noise receiver system.
A first data taking run with 24 days of data does not show evidence for dark photon dark matter in this mass range, excluding dark photon -- photon mixing parameters $\chi \gtrsim 10^{-12}$ in this range at 90\,\% confidence level. This
surpasses existing constraints by about two orders of
magnitude and is the most stringent bound on dark photons in this range below~49\,$\mu$eV.
\end{abstract}


\maketitle

\begin{figure}
	\centering
	\includegraphics[width=\linewidth]{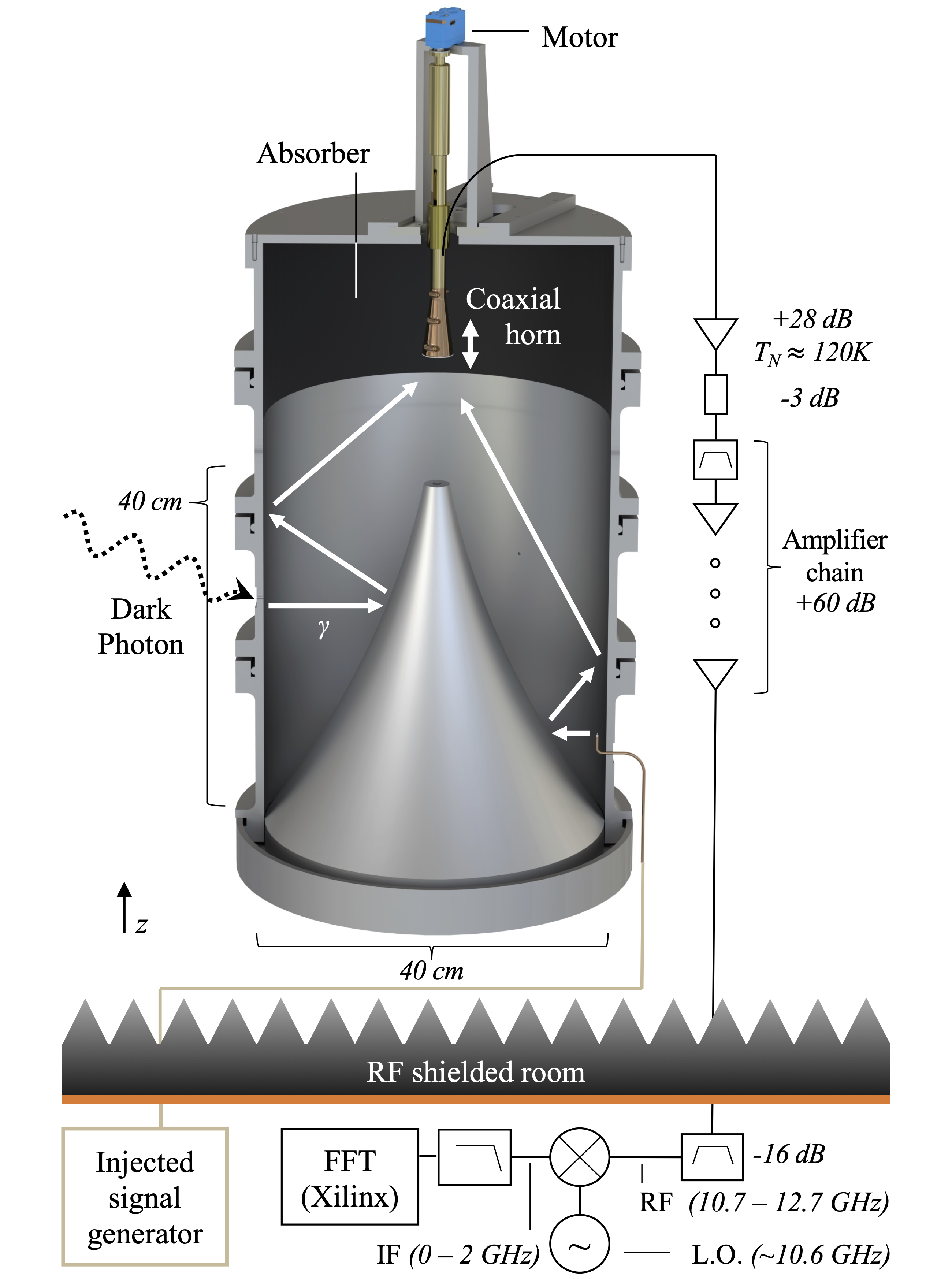}
	\caption{\label{fig:setup} Experimental Setup. DPs convert to photons emitted perpendicularly from the cylinder. The signal is focused on a coaxial horn antenna, amplified using a low-noise receiver chain (right), down-converted and digitized using a custom real-time field-programmable gate array based broadband DAQ (bottom).}
\end{figure}
\emph{Introduction.}
Dark matter (DM) 
remains one of the most elusive enigmas of modern physics~\cite{Rubin:1970zza,Tyson:1998vp,Tegmark:2003ud,Clowe:2006eq,Akrami:2018vks,Bertone:2004pz}. 
Bosonic particles with masses below $\sim \si{\electronvolt}$ (wavelike DM) are well motivated~\cite{Arvanitaki:2009fg,Jaeckel:2010ni,Arias:2012az,Essig:2013lka,Baker:2013zta,Graham:2015rva,Battaglieri:2017aum,Ahmed:2018oog,Irastorza:2018dyq}, and 
many experimental efforts aim to detect them~\cite{Irastorza:2018dyq,RevModPhys.93.015004,Adams:2022pbo}.
Notably the resonant cavity experiments ADMX~\cite{ADMX:2021nhd} and CAPP-12TB~\cite{Yi:2022fmn}
have reached the highly motivated Dine-Fischler-Srednicki-Zhitnitsky threshold for 
QCD axions around $2 - 5\,\mu$eV. Yet, many orders of magnitude in mass range are still unprobed due to the narrowband sensitivity and poor higher-mass scalability of resonant cavities.
Wavelike DM such as dark photons (DPs) can convert to photons emitted perpendicularly to a metallic surface (dish antenna), which can then be focused onto a detector~\cite{Horns:2012jf}. For a surface much larger than the photon wavelength and negligible roughness, this conversion happens independent of mass, overcoming the narrowband limitation of resonant setups.
The signal power is
\begin{equation}
\label{eq:power}
\begin{aligned}
	P\,=&\ \SI{3e-21}{\watt}\,\bigg( \frac{\eta}{0.5} \bigg) \bigg( \frac{A}{\SI{0.5}{\metre\squared} } \bigg)  \\ &\times \bigg( \frac{\chi}{\num{e-12} } \bigg)^2 \bigg( \frac{\rho_{\rm DM}}{\SI{0.45}{\giga\electronvolt\per\centi\metre\cubed} } \bigg) \bigg( \frac{\alpha^2}{{1/3} } \bigg)
\end{aligned}
\end{equation}
at a frequency, $f$, set by the DP mass, $m_{\rm DP}$, and a narrow line shape due to the nonrelativistic DM velocity distribution, $f = m_{\rm DP} c^2 / h + \mathcal{O}(10^{-6})$~\cite{Turner:1990qx,Knirck:2018knd}. $\eta$~is a detection efficiency, $A$~is the emission area of the dish, $\chi$~is the kinetic mixing between photons and DPs, $\rho_{\rm DM}$~is the local DM density and $\alpha$~accounts for the DP's polarization with $\alpha = \sqrt{1/3}$ for an experiment sensitive to a single polarization and randomly polarized~DM.

Several prototype dish antennas have recently been demonstrated~\cite{Suzuki:2015sza,Brun:2019kak,Tomita:2020usq,Ramanathan:2022egk,DOSUE-RR:2022ise,Bajjali:2023uis,Adachi:2023wuo}, but are difficult to combine with high-field magnets to make them sensitive to axion DM.
We recently proposed BREAD (Broadband Reflector Experiment for Axion Detection)~\cite{BREAD:2021tpx}, a dish antenna with a novel focusing reflector geometry shown in figure~\ref{fig:setup}. It allows placement of the emission surface parallel to the field of a solenoid magnet in a future upgrade. In this Letter we present first results from a room temperature pilot experiment, {GigaBREAD}, to search for DPs in the range from $44 - 52\,\mu$eV ($10.7 - \SI{12.5}{\giga\hertz}$).

\emph{Experimental Setup.}
\begin{figure*}
    \centering
    \includegraphics[width=\linewidth]{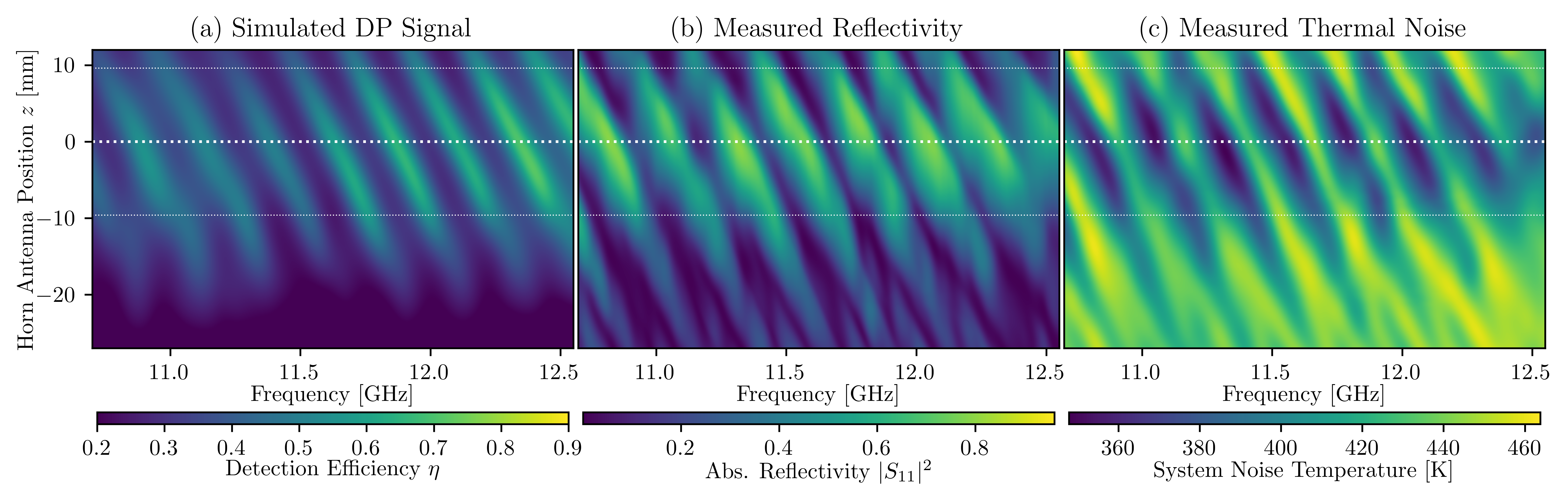}
    \caption{Focal spot measurements vs.\ vertical horn antenna position $z$ and frequency. (a) Expected DP signal from a full-wave Comsol\textsuperscript{\textregistered} simulation. (b) Measured reflectivity. (c) Measured system noise temperature. The focus is symmetric around $z=0$ (bold dotted line). Standing wave resonances between barrel and horn antenna can be swept over frequency by adjusting the horn antenna position between $z \approx \SI{+10}{\milli\metre}$ and $z \approx \SI{-10}{\milli\metre}$ (faint dotted lines).}
    \label{fig:characterization}
\end{figure*}
Figure~\ref{fig:setup} shows a schematic of the experimental setup.
DPs convert to photons emitted perpendicularly from the inner area of the cylindrical aluminum barrel.
They are focused by an inner parabolic reflector and the outer cylinder onto a focal spot at the vertex of the parabolic shape where a coaxial horn antenna~\cite{Barros:2013coax,Bykov:2008coax} is placed. This custom-designed horn receives radiation impinging from the side with radially outward-pointing polarization matched to our geometry, unlike conventional horns typically matched to linearly polarized radiation impinging head on. The horn can be moved vertically in and out of the focal spot.
It is connected to a low-noise amplifier chain with a  heterodyne real-time data acquisition (DAQ) scheme. 
A~pin antenna inserted in front of the conversion surface allows us to inject test signals. 
The dish antenna (cylindrical barrel with reflector and horn antenna), and the  amplifier chain are placed in an RF-shielded Faraday cage with additional foam absorbers. 

The signal from the dish antenna has been simulated with full-wave azimuthally symmetric Comsol\textsuperscript{\textregistered}~\cite{COMSOL} simulations of the modified Maxwell's equations accounting for the wavelike DM field using a space-filling current density~\cite{Knirck:2019eug,Jeong:2023bqb}. 
The simulations include surface deformations of the inner reflector and outer barrel at the level of a few 100\,$\mu$m measured with a coordinate measuring machine.
They shift the focus by a few millimeters axially with otherwise little impact on sensitivity.
Figure~\ref{fig:characterization}\,(a) shows the simulated received power compared to an ideal dish of same size, $\eta$, vs.\ antenna position and frequency of a putative DM signal. The signal vanishes when moving the antenna by about a wavelength out of focus. 
It is weakly resonantly enhanced on a comb of frequencies corresponding to standing waves between the antenna and barrel. Moving the antenna vertically adjusts the corresponding optical length and shifts these frequencies.
Measurements of the reflectivity seen with the horn antenna, fig.~\ref{fig:characterization}\,(b), and the thermal noise, fig.~\ref{fig:characterization}\,(c), reveal the same features and match corresponding simulations.
When measuring reflectivity, the signal emitted from the antenna couples back into the antenna when the antenna is at the focal spot. However, when the antenna is moved out of focus the signal is refocused away from the antenna. Hence, the reflectivity is maximized on focus.
Conversely, thermal radiation from the setup is minimized on focus, because the absorption coefficient equals the thermal emission coefficient according to Kirchhoff's law, i.e., thermal radiation from the absorbers mostly couples into the antenna off focus.
On-resonance photons bounce between the antenna and emission surface, making them more likely to be lost before recoupling to the antenna, leading to dips in the reflectivity and higher thermal noise.
For clarity we removed residual standing waves and small losses from the cable between the horn antenna and amplifier in fig.~\ref{fig:characterization}. These are taken into account in our analysis.

The gain and added noise from the low-noise receiver chain were characterized with the $y$-factor method~\cite{y_factor} using a noise source with \SI{6}{\decibel} excess noise ratio.
The added noise from amplification is \SI[separate-uncertainty = true]{120(10)}{\kelvin} compatible with the low-noise amplifier's data sheet~\cite{lnf_datasheet}. 

After about \SI{80}{\decibel} of amplification the radio frequency (RF) signal passes a $10.7 - \SI{12.7}{\giga\hertz}$ bandpass filter and is down-converted to a so-called intermediate frequency (IF), $f_{\rm IF} = |f_{\rm RF} - f_{\rm LO}| < \SI{2}{\giga\hertz}$,  using a broadband mixer and local oscillator (LO) frequency of $f_{\rm LO} \approx \SI{10.6}{\giga\hertz}$. The readout system is based on the Fermilab Open Source platform QICK~\cite{Stefanazzi:2021otz}. The signal is digitized using a Xilinx RFSoC field-programmable gate array~\cite{xlinlix} (DAQ board)  with an ADC (analog-to-digital converter) at sampling rate \SI{4}{\giga\hertz}.
Custom firmware on the DAQ board Fourier transforms and squares the digitized voltages to obtain power spectra, and finally adds them to previous data in real time.
After 
$10,000$ spectra this sum is transferred to the Pynq~\cite{pynq} logic on the board where the data is stored to an internal hard drive.
The resolution is $\Delta f = \SI{7.8}{\kilo\hertz}$, approximately matching the expected signal line width. 
This choice minimizes the noise competing with a potential signal, while still allowing us to digitize over the maximum bandwidth available with the board.
Low-frequency radio interference (RFI) backgrounds in the IF band can pose a significant challenge for such a DAQ, since it can mimic a DP signal.
To avoid this, we employed a custom frequency hopping scheme: the LO frequency is randomly shifted by an integer number of frequency bins within a few~\si{\mega\hertz} ($\sim 1,000$ bins) about once every second. When averaging the acquired power spectra are accordingly shifted relative to each other in RF frequency domain. This causes any RFI in the IF band to spread across multiple RF frequency bins and average out after sufficiently many acquisitions without affecting RF signals. 
In addition, eight high-power single bin RFI signals were identified in the IF band and digitally masked. 
The frequency hopping also causes these masks to smear out in the RF domain such that their impact on sensitivity becomes negligible.


\emph{Data Taking.} The experiment was operated at 41° 47' 31.6098" N, 87° 36' 6.141” W in the basement of the William Eckhardt Research Center at the University of Chicago.
It is sensitive to DP polarizations perpendicular to Earth's surface, since the focal spot antenna only couples to vertically polarized radiation emitted from the barrel.
During data taking the horn antenna position was regularly swept from $\SI{1}{\centi\metre}$ to $\SI{-1}{\centi\metre}$ around the focus (as indicated in fig.~\ref{fig:characterization}) with a \SI{0.2}{\milli\metre} step size. This tunes the comb of resonances over the full frequency range and gives approximately uniform sensitivity in DP mass.
Each sweep took $\sim 4$ hours, much longer than the frequency hopping intervals, but much shorter than a day to be clearly distinguishable from possible daily modulations~\cite{Knirck:2018knd,Caputo:2021eaa}.
Data taking took place from June 16 to July~17, 2023. It was interrupted deliberately around once per week, in addition to three pauses due to power outages, leading to 24 days of science data. The interruptions were used to remeasure reflectivity as a function of antenna position and to monitor receiver gain by measuring noise from a matched load.
No significant changes were observed except slow gain variations $< \SI{0.1}{\decibel\per{\rm day}}$.
Test signals were injected from June 16 -- June 20 at \SI{11.45}{\giga\hertz} and {July~14} {-- July~17} at \SI{11.53}{\giga\hertz}. 
The injections were blind, i.e., the person first analyzing the data did not know the set~frequency.

\begin{figure*}
    \centering
    \includegraphics[width=\linewidth]{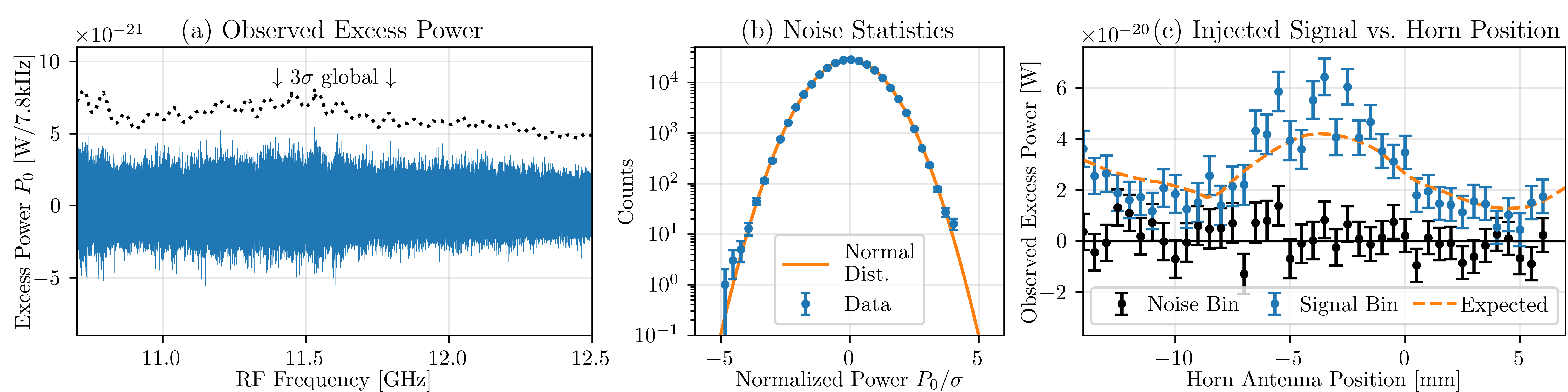}
    \caption{(a) Observed Excess Power $P_0$ vs.\ RF frequency, normalized with receiver gain and detection efficiency. No signal exceeds a global significance of 3$\sigma$. (b) Distribution of the observed excess powers $P_0$ divided by the expected standard deviation from thermal noise $\sigma$, following a normal distribution. (c) Observed excess power for the test signal injected between June 16 and June 20 vs.\ horn antenna position showing the expected weak resonant enhancement.}
    \label{fig:data}
\end{figure*}
\emph{Data Analysis.}
The data were analyzed using standard procedures analogous to other haloscope experiments~\cite{ADMX:2020hay}. 
All acquired power spectra were averaged for each  axial horn antenna position $z$, giving $P_z(f)$. 
We searched for a power excess in a single bin.
To this end, we subtracted the baseline $P_{{\rm bl},z}(f)$ obtained with a fourth-order Savitzky-Golay filter~\cite{doi:10.1021/ac60214a047},
removing features on scales $\gtrsim \SI{1}{\mega\hertz}$ much larger than the signal line width.
The expected standard deviation of residual noise around the baseline is $\sigma_z = P_{\rm bl}/\sqrt{N_{\rm av}}$, where $N_{\rm av}$ is the total number of averages taken at this position. The baseline is related to system noise temperature $T_{\rm sys}$ as $P_{\rm bl} = k_B T_{\rm sys} \Delta f$ with Boltzmann's constant $k_B$ and resolution $\Delta f$. 
The vertical average along $z$ was calculated using an optimally weighted average~\cite{ADMX:2001dbg, Brubaker:2017rna},
\begin{equation}
    P_0 ~ \equiv ~ \left\langle\frac{P}{\eta}\right\rangle_z = \frac{ \sum_z  \left( {\eta_z}/{\sigma_z} \right)^2 P_z / \eta_z }{\sum_z  \left( {\eta_z}/{\sigma_z} \right)^2 }.
\end{equation}
After applying this procedure to the subruns with injected signals, both were identified at the correct frequencies with $30\sigma$ and $40\sigma$ significance, respectively, and bins with injected signals subsequently removed from the analysis.
Fig.~\ref{fig:data}\,(a) shows the obtained excess noise $P_0$ for the full dataset. In absence of a signal $P_0$ is expected to be normal distributed with variance $\sigma^2 = 1/\sum_z (\eta_z/\sigma_z)^2$.
In fig.~\ref{fig:data}\,(b) we histogram excess powers normalized to the expected standard deviation. They are normal distributed, verifying that long-term averaging and background subtraction work as expected.
Fig.~\ref{fig:data}\,(c) illustrates the dependence of the test signal injected from June 16 to Jun 20 on the vertical antenna position showing the expected weak resonant enhancement around a certain antenna position.
   
Signals are searched using a cross-correlation of $P_0$ with the expected Maxwell-Boltzmann line shape~\cite{Turner:1990qx}.
The largest excess is observed at a local significance of $5.2\sigma$ at \SI{10.829}{\giga\hertz}. Its global significance is less than $2\sigma$ as estimated using Monte Carlo simulations, too small to claim a detection.
We derive a 90\,\% confidence level (CL) upper limit on the received DM power based on the observed excess powers greater than zero.
Applying eq.~\ref{eq:power} gives the limit on the mixing parameter $\chi$.

\begin{table}[]
\caption{\label{tab:systematics} Frequency-averaged systematic uncertainties on the DP signal power.
}
\vspace{0.2cm}
\begin{tabular}{@{}lr@{}}
\toprule
Effect                                  & Uncertainty on $P_{0}$ \\ \midrule
Nonzero DM velocity & $< \SI{1}{\percent}$\\
Detection Efficiency Simulation  & \SI{20}{\percent}\\[2pt]
System Noise Temperature & \SI{9}{\percent}  \\
Gain Variations & \SI{5}{\percent} \\[2pt] 
Baseline Removal & \SI{3}{\percent} \\
\midrule
Total & \SI{23}{\percent}\\
\bottomrule
\end{tabular}
\end{table}

Table~\ref{tab:systematics} summarizes systematic uncertainties.
Momentum-transfer from DM to the converted photons can shift the focus~\cite{Jaeckel:2013sqa,Jaeckel:2015kea,Jaeckel:2017sjb}. This effect is negligible since our setup is much smaller than the DP de~Broglie wavelength $\lambda_{\rm DB} \sim \SI{30}{\metre}$.
The largest systematic arises from uncertainties in the simulated efficiency $\eta$. It is estimated based on the differences in the measured and simulated reflectivity $S_{11}$ and the corresponding impact on signal power.
The differences may be attributed to geometrical imperfections not taken into account in the simulation.
Another important systematic arises from the $y$-factor calibration of the low-noise amplifier's added noise, besides small gain drifts discussed above.
The baseline removal in the analysis can attenuate signals~\cite{Brubaker:2017rna,Diehl:2023fuk}. This effect is smaller than \SI{3}{\percent} for our filter parameters.
The total systematic uncertainty is obtained by adding the individual uncertainties in quadrature. The most conservative $P_0$ within this uncertainty is adopted for the limit.

\begin{figure*}
    \centering
    \includegraphics[width=\linewidth]{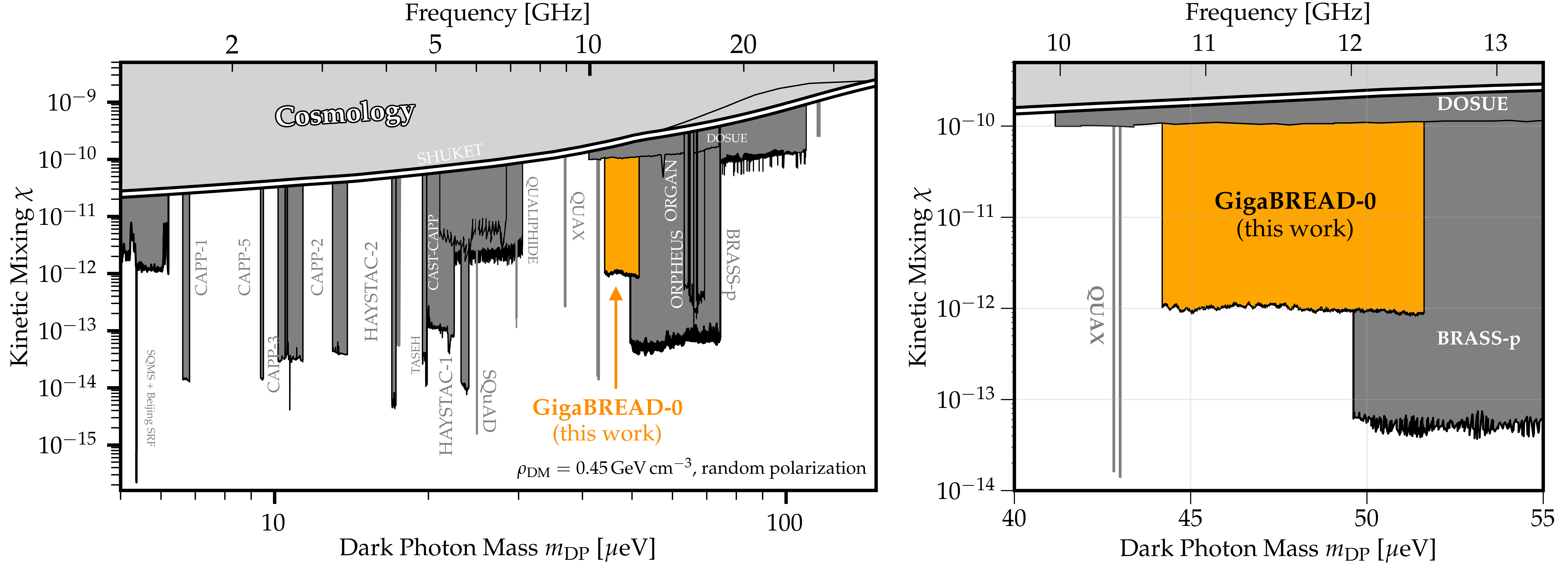}
    \caption{DP parameter space excluded in this work (GigaBREAD-0), compared to cosmological constraints, other dish antennas like SHUKET~\cite{Brun:2019kak}, QUALIPHIDE~\cite{Ramanathan:2022egk}, BRASS-p~\cite{Bajjali:2023uis} and 
    DOSUE~\cite{DOSUE-RR:2022ise,Adachi:2023wuo}, and other haloscopes~\cite{AxionLimits} over about two decades in mass (left) and magnified (right). }
    \label{fig:limit}
\end{figure*}
\emph{Conclusion.} We obtain an upper bound of $\chi \lesssim 10^{-12}$ for masses of $44 - 52\,\mu$eV at 90\,\% CL as shown in 
fig.~\ref{fig:limit}. This surpasses cosmological bounds by more than two orders of magnitude, and is the most sensitive experimental constraint between $44$ and $49\,\mu$eV.
Furthermore, our setup is optimized for inclusion in a solenoid magnet and a first run in a \SI{4}{\tesla} field~\cite{4t_magnet} is under preparation to search for axion-like particles.
Our setup establishes a testbed for future research and development directions, e.g., cryogenic quantum-limited amplifiers~\cite{Ramanathan:2022egk}, other photo-sensing technologies and larger dish sizes. This also includes signal enhancement by improving quality factors of the standing wave resonances or larger volume efficiency with dielectric stacks~\cite{Caldwell:2016dcw,Millar:2016cjp,Chiles:2021gxk,Manenti:2021whp}.

\begin{acknowledgments}
\emph{Acknowledgments.} We acknowledge Abigail Vieregg and her group for providing access to the RF shielded room.
We thank Dan Zhang and the ADMX Collaboration for helpful discussions about analysis. 
We thank all BREAD collaborators for inspiring and helpful discussions.
This work is funded by the by the Department of Energy through the resources of the Fermi National Accelerator Laboratory (Fermilab), a U.S. Department of Energy, Office of Science, HEP User Facility. Fermilab is managed by Fermi Research Alliance, LLC (FRA), acting under Contract No. DE-AC02-07CH11359.
J.\ L.\ is supported by a Junior Research Fellowship at Trinity College, University of Cambridge.
C.\ P.\ S.\ is supported by the Kavli Institute for Particle Astrophysics and Cosmology Porat Fellowship.
\end{acknowledgments}

\end{document}